\documentclass[sigconf,edbt,dvipsnames]{acmart-edbt2019}

\usepackage{comment}
\usepackage{enumerate}
\usepackage{color}
\usepackage{relsize}
\usepackage{algorithm}
\usepackage{algpseudocode}

\setcopyright{rightsretained}

\acmDOI{}

\acmISBN{XXX-X-XXXXX-XXX-X}

\begin{document}

\title{Stateless and Non-Interactive Order-Preserving Encryption for Outsourced Databases through Subtractive Homomorphism}






\author{Dongfang Zhao
\\
	\textit{University of Washington}
\\dzhao@uw.edu
}

\begin{abstract}
Order-preserving encryption (OPE) has been extensively studied for more than two decades in the context of outsourced databases because OPE is a key enabling technique to allow the outsourced database servers to sort encrypted tuples in order to build indexes, complete range queries, and so forth.
The state-of-the-art OPE schemes require (i) a stateful client---implying that the client manages the local storage of some mapping between plaintexts and ciphertexts, and/or (ii) the interaction between the client and the server during the query.
In production systems, however, the above assumptions do not always hold (not to mention performance overhead):
In the first case, the storage requirement could exceed the capability of the client;
In the second case, the clients may not be accessible when the server executes a query involving sort or comparison.

This paper proposes a new OPE scheme that works for stateless clients and requires no client-server interaction during the queries.
The key idea of our proposed protocol is to leverage the underlying additive property of a homomorphic encryption scheme such that the sign of the difference between two plaintexts can be revealed by some algebraic operations with an evaluation key.
We will demonstrate the correctness and security of the proposed protocol in this short paper;
the implementation and experimental results will be presented in an extended report.
\end{abstract}

\maketitle

\section{Introduction}

\subsection{Background and Motivation}

For the last two decades, we have witnessed the inception and prosperity of database as a service (DaaS) since the seminal paper~\cite{haci_icde02} in ICDE'02.
As of the writing of this paper, all major cloud computing vendors (e.g., Amazon Web Services, Google Cloud Platform, Microsoft Azure) support DaaS,
which allows users to avoid the upfront cost of managing their in-house databases.
As with any outsourced service, the security of outsourced databases has been one of the top concerns for users;
One avenue of research to address the security concern is to encrypt the user's sensitive data before uploading them to the outsourced database~\cite{popa2011cryptdb}.

As a double-edged sword, encrypted tuples bring up new challenges in outsourced databases:
Critical workloads like building indexes cannot be supported in the database system due to the randomness of encrypted tuples.
To that end, in SIGMOD'04, Agrawal et al.~\cite{rag_sigmod04} proposed to encode the plaintext in the outsourced database while retaining the numerical order of the corresponding plaintexts;
however, the approach generated deterministic ciphertexts that can leak the plaintext repetition.


In EuroCrypt'09, Boldyreva et al.~\cite{abold_eurocrypt09} presented the first security definition of OPE.
Following the convention of cryptography, the definition is based on the canonical structure of an encryption scheme:
(i) The security goal is computational indistinguishability,
(ii) The threat model is to allow the adversary to obtain a polynomial number of ciphertexts of arbitrary plaintexts,
i.e., the so-called chosen-plaintext attack (CPA), and
(iii) A simulation-based reduction is suggested to prove that the distinguishability between two ciphertexts is negligible.
Unfortunately, it was shown that the indistinguishability under the standard CPA model was impossible because the CPA definition was overly strong and could be easily violated if the adversary learned about the orders of the plaintexts.
To that end, a relaxed security notion,
namely \textit{indistinguishability under ordered chosen-plaintext attack} (IND-OCPA),
was proposed by~\cite{abold_eurocrypt09}.
IND-OCPA was similar to IND-CPA except for allowing the adversary to learn about the order of the plaintext.

Not long after IND-OCPA was proposed, 
Popa et al.~\cite{rpopa_sp13} in Oakland'13 pointed out that IND-OCPA was vulnerable to the well-known frequency attack.
As a result, a stronger security notion~\cite{fker_ccs15} was proposed in CCS'15,
i.e., \textit{indistinguishability under frequency-analyzing and ordered chosen-plaintext attack} (IND-FAOCPA).
Multiple ``modern'' OPE schemes claimed to meet IND-FAOCPA,
such as~\cite{droche_ccs16,dli_vldb21};
however, one recent work~\cite{xcao_vldb23} in VLDB'23 pointed out that there was an inherent barrier to support fully IND-FAOCPA if checking ciphertext equality prohibited.

While the cryptography and security communities strive to define and prove the OPE security from a theoretical perspective,
the database and system communities are equally interested in developing practical systems and services based on OPE,
such as Microsoft Azure~\cite{panto_sigmod20} as reported in SIGMOD'20.
An evaluation paper~\cite{dboga_vldb19} in VLDB'19 provided a thorough summary of major OPE schemes as of 2019.

As of the writing of this paper, the latest OPE scheme in outsourced databases remains OPEA proposed in VLDB'21~\cite{dli_vldb21}.
The limitation of OPEA, as well as many other OPE schemes, 
lies at the client storage.
As a stateful scheme, OPEA requires the client to maintain a local table to keep track of the plaintext orders.
Some alternative solutions, 
such as POPE~\cite{droche_ccs16},
do not suffer from the client storage,
and yet have other limitations,such as a non-constant number of communication rounds between the client and the server during the queries.
\textit{In summary, state-of-the-art OPE solutions in outsourced databases suffer from the storage overhead on stateful clients and/or the communication overhead between the client and the server during the queries.}

\subsection{Proposed Work}

In this work, we propose a new OPE scheme, called Homomorphic OPE (HOPE), which 
(i) incurs zero client storage, making the client completely stateless; and
(ii) incurs zero communication between the client and the server, allowing the outsourced database server to reveal the order of encrypted tuples offline.
The key idea of HOPE is to leverage the additive property of homomorphic encryption such that the comparison between two encrypted tuples can be transformed into a randomized difference between the two tuples while retaining the sign between their subtraction.
Specifically, we will present the following in the remainder of this short paper:
\begin{itemize}
    \item A new OPE scheme called HOPE is designed that does not incur any client storage or client-server communication;~(Section~\S\ref{sec:hope})
    \item The security of HOPE is formally proved.~(Section~\S\ref{sec:security})
\end{itemize}

\section{Related Work}

\subsection{Order-Preserving Encryption in Databases}

The concept of order-preserving encryption (OPE) was originally proposed in the database community~\cite{rag_sigmod04}.
The motivation is evident:
how could we achieve both the confidentiality and the orders of sensitive data in an outsourced database?
The confidentiality part is obvious and the order requirement is also well justified:
it is very common for database systems to build $B^+$-tree indexes to speed up the range and insertion queries and being able to sort or order the outsourced data sets is essential to achieve this goal.

The conventional solution to achieve the dual goals is somewhat straightforward:
the plaintexts are encoded with the help of some statistical distribution such that the encoded values remain in the same order as the plaintext,
which was insecure because the encoded values are deterministic~\cite{panto_sigmod20}.
The solution is to introduce some function for the database to order the encrypted tuples without relying on the raw values of ciphertexts,
which is called \textit{order-revealing encryption} (ORE)~\cite{dbone_eurocrypt15}.
Accordingly, a new security notion was proposed to allow the adversary to learn about the orders of plaintexts,
resulting in the so-called \textit{indistinguishability under ordered chosen-plaintext attack} (IND-OCPA).
Although there are many methods to calculate the order values,
such methods can be categorized into two types:
(i) a stateful scheme where the client and the server coordinate to maintain the order information of encrypted records in the database~\cite{rpopa_sp13}, and
(ii) a stateless scheme where the order information can be retrieved on the fly~\cite {nshen_prdc21}.
Most OPE schemes in the literature focus on the stateful approach;
The proposed HOPE scheme is stateless to avoid the storage and performance overhead.

It turned out that there were new issues for ORE and IND-OCPA:
Multiple IND-OCPA schemes~\cite{abold_eurocrypt09,rpopa_sp13} are vulnerable to attacks that leverage the access patterns of the queries.
To that end, a newer notion is defined,
namely \textit{indistinguishability under frequency-analyzing ordered chosen-plaintext attack} (IND-FAOCPA).
Multiple IND-FAOCPA schemes have been proposed in the literature,
such as~\cite{dli_vldb21,fker_ccs15,droche_ccs16}.
A relatively recent evaluation paper~\cite{dboga_vldb19} reports the performance of some of the most popular OPE schemes,
including the aforementioned schemes and a few others~\cite{nchen_fse16,klewi_ccs16,dcash_asiacrypt}.
It was reported that leveraging homomorphic encryption~\cite{ppail_eurocrypt99} and garbled circuit~\cite{szahu_eurocrypt15} could further improve the query performance of OPE~\cite{atueno_accs20};
however, it is unclear how to support efficient insert operations.
As of the writing of this paper, OPEA~\cite{dli_vldb21} achieves the best performance in almost all the metrics,
although we are not aware of any production database systems taking this approach.

\subsection{Homomorphic Encryption in Databases}

The notion \textit{homomorphism} refers to a class of functions that preserve the algebraic structures of the input and output spaces.
More specifically, an algebraic group\footnote{We use \textit{algebraic group} to refer to a group structure in group theory, \textit{not} the solutions to a system of polynomial equations in algebraic geometry.} can be relabelled and transformed, through a homomorphic function, into another algebraic group without changing the relationship among the elements. 
An algebraic \textit{group} is defined a nonempty set of elements along with a binary operator satisfying the closure, associativity, identity, and inverse properties.


\textit{Homomorphic encryption} (HE) is a specific type of encryption where certain algebraic operations between operands in the plaintext space (e.g., group $G$) can be semantically mapped to well-defined functions over the elements in the ciphertext space (e.g., group $H$).
For example, if an HE encryption algorithm $Henc(\cdot)$ is additive,
then the plaintexts with $+$ operations can be translated into a homomorphic addition $\oplus$ on the ciphertexts.
Formally, if $a$ and $b$ are plaintexts, the relationship is defined as:
\[
Dec(Henc(a) \oplus Henc(b)) = a + b,
\]
where $Dec$ denotes the decryption algorithm.
As a concrete example, setting $Henc(x) = 2^x$ (temporarily disregarding security considerations of $Henc(\cdot)$) demonstrates that $Henc(a+b) = 2^{a+b} = 2^a \times 2^b = Henc(a) \times Henc(b)$, indicating that $\oplus$ corresponds to arithmetic multiplication $\times$.

An HE scheme enabling additive operations is termed \textit{additive}.
Popular additive HE schemes include Paillier~\cite{ppail_eurocrypt99},
which is an asymmetric scheme using a pair of public and private keys. 
An HE scheme that supports multiplication is said \textit{multiplicative}.
Similarly, a multiplicative HE scheme guarantees the following equality,
\[
Dec (Henc(a) \otimes Henc(b)) = a \times b,
\]
where $\otimes$ denotes the homomorphic multiplication over the ciphertexts.

An HE scheme that supports both addition and multiplication is called a \textit{Fully Homomorphic Encryption (FHE)} scheme.
This requirement should not be confused with specific addition and multiplication parameters, such as Symmetria~\cite{symmetria_vldb20} and NTRU~\cite{ntru}.
That is, the addition and multiplication must be supported homomorphically under the same scheme $Henc(\cdot)$:
\[\displaystyle
\begin{cases}
    Dec( Henc(a) \oplus Henc(b) ) = a + b, \\
    Dec( Henc(a) \otimes Henc(b) ) = a \times b.
\end{cases}
\]
Constructing FHE schemes remained a formidable challenge until Gentry~\cite{cgentry_stoc09} presented a feasible approach using lattice ideals.
Subsequent generations of FHE schemes,
e.g., \cite{bgv,bfv,ckks}, had exhibited substantial improvements in encryption efficiency, partially due to the removal of ideal lattices;
Instead of using ideal lattices, those newer FHE schemes are based on the learning with error (LWE)~\cite{oregev_jacm09} or its variant ring learning with error (RLWE),
which have been proven to be as hard as hard lattice problems (e.g., through quantum or classical reduction).
Open-source libraries of FHE schemes, such as IBM HElib~\cite{helib} and Microsoft SEAL~\cite{sealcrypto}, are available.

Applying HE schemes to outsourced database has recently drawn interests in the dataase community.
Symmetria~\cite{symmetria_vldb20} is a recent scheme proposed in the database community,
which is multiplicative using a distinct scheme from the one for addition.
Some recent advances in applying HE to database systems can be found in~\cite{otawose_sigmod23}, where both caching and parallel processing are proposed to accelerate the HE procedure in typical database workloads.
Other schemes with the HE property include RSA~\cite{rsa} and ElGamal~\cite{elgamal_tit85}, serving as candidate schemes for HE in databases.

\section{Homomorphic Order-Preserving Encryption}
\label{sec:hope}

\subsection{Notations and Preliminaries}
\label{sec:sub_notation}

\paragraph{Number theory and abstract algebra}
We use $\mathbb{Z}$ to indicate the integer set;
$\mathbb{Z}^+$ denotes the positive integer set and $\mathbb{Z}_n$ is the set of integers between 0 and $n-1$, where $n \in \mathbb{Z}^+$.
Unless otherwise stated, $p$ and $q$ denotes two distinct odd prime integers (thus $p,q > 2$).
If $x$ does not divides $y$, we say $x \nmid y$.
We call two distinct positive integers \textit{co-prime} if their greatest common divisor (GCD) is 1, or $GCD(\cdot, \cdot) = 1$.
The number of co-prime integers with $n$ is denoted by a function $\varphi(n)$.
The set of co-prime integers with $n$ is denoted by $\mathbb{Z}^*_n$; 
hence, by definition the cardinality of $\mathbb{Z}^*_n$ is $\phi(n)$, or $\left| \mathbb{Z}^*_n \right| = \phi(n)$.
The set $\mathbb{Z}^*_n$ of integers along with the arithmetic multiplication modulo $n$, denoted by $\times_n$, forms a (multiplicative) \textit{group} $(\mathbb{Z}^*_n, \times_n)$,
which means (i) the operation $\times_n$ is closed and associative, (ii) the set $\mathbb{Z}^*_n$ has an identity element, and (iii) every element has an inverse element in $\mathbb{Z}^*_n$.
If the context is clear, we may use $\mathbb{Z}^*_n$ to denote the group $(\mathbb{Z}^*_n, \times_n)$.
The order of a group is defined as the cardinality of its set.
Evidently, $\left| \mathbb{Z}^*_p \right| = p - 1$;
without much effort, we can show that $\left| \mathbb{Z}^*_{pq} \right| = (p-1)(q-1)$ and $\left| \mathbb{Z}^*_{p^2q^2} \right| = p(p-1)q(q-1)$.
One well-known result\footnote{Which is also referred to as the Fermat's little theorem in a more specific scenario.} in (finite) group theory is that any element after being raised to the order of the group will be equal to the identity element.
For example, let $r$ be a uniformly sampled element from a specific set, denoted by $r \gets \mathbb{Z}^*_{p^2q^2}$,
then the following equality holds
\[
r^{p(p-1)q(q-1)} \equiv 1 \text{ (mod } p^2q^2 \text{)}.
\]
Given any element $e$ in a group $G$, we can efficiently compute the the inverse (denoted by $e^{-1}$) of $e$ using the well-known Extended Euclidean Algorithm.

\paragraph{Cryptographic primitives}
By convention, we use $Enc()$ and $Dec()$ to denote the encryption and decryption algorithms of a scheme.
Unless otherwise stated, the key generation algorithm $Gen()$ always takes in a string of $\lambda$ bits, denoted by $1^{\lambda}$.
In homomorphic encryption schemes, there are usually one or multiple \textit{evaluation keys} (EKs) that allow the untrustful parties to carry out certain algebraic operations, 
such as plus, multiplication, and specifically comparison in this work.
For example, to support arithmetic addition over ciphertexts, the evaluation function $EvalAdd()$ might look like the following
\[
c_{add} = EvalAdd_{ek}(c_0, c_1),
\]
such that
\[
Dec(c_{add}) = Dec(c_0) + Dec(c_1),
\]
where $ek$ denotes the evaluation key and $c_i \gets Enc(m_i)$ for plaintext messages $m_i$, $i \in \{0, 1\}$.
Similarly, let $Sgn()$ denote the sign function, then the homomorphic sign function, $EvalSign()$, satisfies the following property
\[
EvalSign\left[ EvalAdd(c_0, \ominus c_1) \right] = Sign(m_0- m_1),
\]
where $\ominus$ denotes the negation operation in the ciphertext space,
which means that we can carry out homomorphic subtraction as well.
A negation/subtraction operation is not always available in (additive) homomorphic encryption schemes; 
the proposed protocols in this work will be built upon those schemes that permit homomorphic subtractions.

\paragraph{Paillier}
HOPE is extended from a well-known \textit{additive} homomorphic encryption scheme Paillier~\cite{ppail_eurocrypt99}.
We briefly review the key algorithms in Paillier in the following.
Let $n = pq$, $p \nmid (q-1)$, and $q \nmid (p-1)$, then we have $GCD(n, \phi(n)) = 1$.
The public key is $n$ and the private key is $\phi(n) = pq$.
The encryption of Paillier is defined as
\[
Enc^{Pail}(m) = (n+1)^m r^n,
\]
where $m\in \mathbb{Z}_n$ is the plaintext and $r$ is randomly sampled from $\mathbb{Z}^*_n$.
The decryption is defined as
\[
Dec^{Pail}(c) = \left( \frac{\left( c^{\phi(n)} \text{ mod } n^2 \right) - 1}{n} \right) \times \left(\phi(n)^{-1}\text{ mod } n \right) \text{ mod } n,
\]
where $c\in \mathbb{Z}^*_{n^2}$ is the ciphertext.
Lastly, the addition between two ciphertexts $c_0$ and $c_1$ are defined as follows:
\[
c_0 \oplus c_1 = EvalAdd^{Pail}(c_0, c_1) = c_0 \times c_1 \text{ mod } n^2,
\]
where $\oplus$ denotes the addition operator in the ciphertext space.
The correctness and security of Paillier have been well studied in the literature and we will not review them here.

\paragraph{Security proof}
As all other OPE schemes, our proposed HOPE protocol aims to achieve \textit{computational security}, meaning that an efficient, or probabilistic polynomial time (PPT), adversary $\mathcal{A}$ could distinguish a ciphertext encrypted by HOPE and a random string with a probability bounded by a \textit{negligible function} even if $\mathcal{A}$ can conduct a comparison query between a polynomial number of ciphertext pairs.
A \textit{negligible function} $\mu(n)$ is a function whose absolute value decreases faster than the inverse of any polynomial functions;
in practice, we usually, but not always, select the inverse of an exponential function to serve as the $\mu(n)$.
We use $Poly(n)$ to denote the class of polynomial functions in $n$;
if the context is clear, we will also use $Poly(n)$ to denote an unspecified polynomial function. 
It can be shown that if $\mu_1(n)$ and $\mu_2(n)$ are both negligible functions, so are $\mu_{add}(n) = \mu_1(n) + \mu_2(n)$ and $\mu_{mul}(n) = \mu_1(n) \times \mu_2(n)$.
There are multiple templates for theoretical security proof, such as reduction and simulation,
although the core ideas are similar:
The newly proposed scheme is compared to a baseline scheme (which could be a well-accepted secure scheme or even a perfectly secure world) and the proof shows that the two schemes are not computationally distinguishable.
Our security proof of HOPE will be based on the popular \textit{reduction} template, in which we will show that being able to break HOPE would entail the insecurity of a proved secure homomorphic encryption, leading to a contradiction.

\subsection{Negative Plaintexts and Subtraction}
\label{sec:sub_smod}

The vanilla version of Paillier works on the plaintext space of $\mathbb{Z}_{pq}$ and does not support negative plaintexts or subtraction.
However, in order to compare two plaintexts, HOPE (and also many other OPE schemes) relies on the sign of the subtraction between the plaintexts.
Therefore, we first describe how to extend Paillier into negative plaintexts and subtraction over the plaintexts.

\paragraph{Handling negative plaintexts}
We need to modify the encryption algorithm in order to support negative plaintexts.
The new encryption is as follows:
\[
Enc^{Hope}(m) = Enc^{Pail}(m \text{ mod } n),
\]
where $m \in \left[-\left\lfloor \frac{n}{2} \right\rfloor, \left\lfloor \frac{n}{2} \right\rfloor \right)$.
Correspondingly, the decryption works as follows:
\[
Dec^{Hope}(c) = \left(Dec^{Pail}(c) + \left\lfloor \frac{n}{2} \right\rfloor \text{ mod } n\right) - \left\lfloor \frac{n}{2} \right\rfloor,
\]
where $c$ is in the same ciphertext space $c \in \mathbb{Z}_{n^2}^*$.
What the above equation essentially does is to shift the value to the $-\infty$ direction by $\left\lfloor \frac{n}{2} \right\rfloor$;
because we will reuse this operation in the remainder of this paper,
we will call such a shifting operation by \textit{symmetric modulo} (smod) and we can rewrite the equation as
\[
Dec^{Hope}(c) = Dec^{Pail}(c) \text{ smod } n.
\]
As a result, the negation of a plaintext can be calculated in the ciphertext space as follows:
\[
\ominus c = EvalNeg^{Hope}(c) = c^{-1} \text{ mod } n^2,
\]
and it is guaranteed that 
\[
Dec^{Hope}(\ominus c) = -m.
\]

\paragraph{Subtraction}
After extending the plaintext space into negative integers, it is fairly nature to support subtractions in the ciphertext space, which we refer to as \textit{subtractive homomorphism}:
\[
c_0 \ominus c_1 = EvalSub^{Hope}(c_0, c_1) = EvalAdd^{Hope}\left( c_0, \ominus c_1 \right),
\]
where $c_i \gets Enc^{Hope}(m_i)$, $i \in \{0, 1\}$.
It is not hard to show that the following equality holds:
\[
Dec^{Hope}\left( c_0 \ominus c_1 \right) = m_0 - m_1.
\]

\subsection{Homomorphic Comparison}

We will introduce a new evaluation function for comparing two ciphertexts in the sense that the sign of the ciphertext subtraction is identical to the sign of the corresponding plaintexts.
That is, let $c_i \gets Enc^{Hope}(m_i)$, $i \in \{0, 1\}$, 
we aim to design a function $EvalSign()$ such that the following equation holds:
\[
EvalSign^{Hope}(c_0 \ominus c_1) = Sgn(m_0 - m_1),
\]
where $Sgn()$ is a function defined as below:
\[
Sgn(x) = 
\begin{cases}
    1, \text{ if } x > 0,\\
    0, \text{ if } x = 0,\\
    -1, \text{ if } x < 0. 
\end{cases}
\]
We can define the homomorphic comparison function based on the above homomorphic sign function as follows:
\[
EvalCmp^{Hope}(c_0, c_1) = EvalSign^{Hope}(c_0 \ominus c_1). 
\]

Many existing OPE schemes~\cite{xcao_vldb23} forbid the $Sgn(0) = 0$ case because it would allow the adversary to launch the frequency analysis, i.e., the adversary could measure the frequency of messages in the application. 
Usually what those scheme did was to randomize the ciphertext-comparison result if the two corresponding plaintexts are equal.
However, in database indexing and range queries, being able to tell the equality among tuples is critical (e.g., \texttt{group by} in SQL).
Moreover, formally speaking, the standard plaintext $Sgn()$ function does reveal the equality between plaintexts.
Therefore, HOPE follows the ternary semantics that allows equality queries.

Per the definition of computational security, $EvalSign^{Hope}()$ should not reveal more information than an $Sgn()$ function does by up to a negligible function.
Therefore, $EvalSign^{Hope}()$ must satisfy the following properties:
(i) it should not reveal any new information about the secret key $\phi(n)$ by up to a negligible function in $n$, and
(ii) it should not change the distribution of the ciphertext space in the view of any PPT adversary.
We will present a thorough analysis of HOPE security in Section~\S\ref{sec:security};
this section will focus on the description of the homomorphic comparison algorithm and its correctness.

\paragraph{Evaluation key}
We start by constructing the evaluation key ($ek$) of HOPE.
Generally speaking, an $ek$ is accessible to the outsourced database such that the latter could compute the output of $EvalSign^{Hope}()$ for building index, executing \texttt{group by} queries, etc.
Because \textit{ek} we will construct is only useful for computing $EvalSign^{Hope}()$ (i.e., $ek$ is not needed for $EvalAdd^{Hope}()$ or $EvalNeg^{Hope}()$), 
we will call this $ek$ a \textit{comparison key} ($ck$).
Let $\zeta, \eta_0, \eta_1 \in \mathbb{Z}^*_n$ be a random integers.
The $ck$ is defined as a pair $ck = (ck_0, ck_1)$:
\[
\begin{cases}
    ck_0 = \eta_0 \cdot \phi(n)^{\zeta} \text{ mod } n\\
    ck_1 = \eta_1 \cdot \left( \phi(n)^{\zeta} \right)^{-1} \text{ mod } n
\end{cases}
\]
The evaluation of homomorphic sign function is then defined as follows:
\[
\begin{split}
& EvalCmp^{Hope}(c_0, c_1) \\
=\quad & \frac{\left(\left(\frac{c_0}{c_1}\right)^{ck_0} \text{ mod } n^2\right) - 1}{n} \times \left( ck_1 \text{ mod } n \right) \text{ smod } n,
\end{split}
\]
where ``smod $n$'' was previously defined as shifting the output toward $-\infty$ by a half of the domain range in~\S\ref{sec:sub_smod}.
For security reasons, $ck$ should be updated periodically, which will be discussed in Section~\S\ref{sec:security}.

\subsection{Correctness}
\label{sec:sub_correctness}

In order to verify the correctness of $EvalCmp^{Hope}()$ function,
we need to observe a few facts. 
Some of them are trivial and in that case we will simply state the results;
others need a bit algebraic work and for them we will briefly sketch the proof.

\paragraph{Computing the comparison key}
We need to verify that the data owner can efficiently compute the ck.
While $ck_0$ can be obviously computed by modulo exponentiation because the owner knows $\phi(n) = (p-1)(q-1)$, 
the feasibility of computing $ck_1$ depends on whether the owner can efficiently compute the inverse of $\phi(n)^\zeta$, if it exists.
To show that the owner could indeed do this,
we need to verify that $\phi(n)^\zeta$ is an element in a multiplicative group.
We claim that $GCD(\phi(n), n) = 1$ because the only factors of $n$ are 1, $p$, $q$, and $n$; 
however, we know that 
(i) $n \nmid \phi(n)$ because $n > \phi(n)$;
(ii) $p \nmid \phi(n)$ because $p \nmid (p-1)$ (i.e., $p > p-1$) and $p \nmid (q-1)$ (by definition in~\S\ref{sec:sub_notation}); and
(iii) $q \nmid \phi(n)$ because $q \nmid (q-1)$ (i.e., $q > q-1$) and $q \nmid (p-1)$ (by definition in~\S\ref{sec:sub_notation}).
Therefore, the only common factor between $\phi(n)$ and $n$ is 1,
which means $\phi(n) \in \mathbb{Z}^*_n$.
This implies that $\phi(n)^\zeta \in \mathbb{Z}^*_n$ by the closure property of the multiplicative group $\mathbb{Z}^*_n$.
As a result, the inverse of $\phi(n)^\zeta$ can be efficiently computed using the extended Euclidean algorithm.
Thus, $ck_1$ can be efficiently computed by one additional modulo $n$.

\paragraph{Homomorphic comparison function}
Before demonstrating the correctness of $EvalCmp^{Hope}()$, 
we need a couple of lemmas.
\begin{lemma}\label{thm:binomial}
    For positive numbers $n > 1$ and $x > 0$, the following equality holds:
    \[
    (1 + n)^x = 1 + nx \text{ mod } n^2.
    \]
\end{lemma}
\begin{proof}
    The claim can be proved using a straightforward polynomial (i.e., binomial) expansion:
    \[
    (1+n)^x = \sum_{i=0}^n \binom{x}{i} 1^{n-i} n^i = 1 + nx + n^2\binom{x}{2} + \dots = 1 + nx \text{ mod } n^2.
    \]
\end{proof}
\begin{lemma}\label{thm:lagrange}
Let $n = pq$, where $p$ and $q$ are distinct primes. Then for any $r \in \mathbb{Z}^*_{n^2}$ the following equation holds:
\[
r^{n\phi(n)} = 1 \text{ mod } n^2.
\]
\end{lemma}
\begin{proof}
    The order of group $\mathbb{Z}^*_{n^2}$ is $\phi(n^2)$, which is equal to $p(p-1)q(q-1) = (pq)\cdot \left[ (p-1)(q-1) \right] = n\phi(n)$.
    It is well known that any element raised to the order of the finite group is equivalent to the identity of the group. 
    Therefore, the following holds:
    \[\displaystyle
    r^{n\phi(n)} = r^{\left|\mathbb{Z}^*_{n^2}\right|} = 1 \text{ mod } \left|\mathbb{Z}^*_{n^2}\right| = 1 \text{ mod } n^2,
    \]
    as desired.
\end{proof}
Assuming $c_i = (1+n)^{m_i}\cdot r_i^n$, $i \in \{0, 1\}$,
we can expand the formula of $EvalCmp^{Hope}()$ as follows:
\[
\begin{split}
& EvalCmp^{Hope}(c_0, c_1) \\
=\quad & \frac{(n+1)^{ck_0(m_0-m_1)}\left(\frac{r_0}{r_1}\right)^{n\cdot ck_0} \text{ mod } n^2 - 1}{n}\\
&\times \left( ck_1 \text{ mod } n \right) \text{ smod } n\\
=\quad & \frac{(1+n)^{\eta_0\phi(n)^\zeta(m_0-m_1)}\left(\frac{r_0}{r_1}\right)^{n\cdot \eta_0\phi(n)^\zeta} \text{ mod } n^2 - 1}{n}\\ 
&\times \left(\eta_1 \cdot \left( \phi(n)^{\zeta} \right)^{-1} \text{ mod } n \right) \text{ smod } n,
\end{split}
\]
Per Lemma~\ref{thm:binomial}, we know that 
\[
(1+n)^{\eta_0\phi(n)^\zeta(m_0-m_1)} = 1 + n\eta_0\phi(n)^\zeta(m_0-m_1) \text{ mod } n^2.
\]
Therefore, the above formula can be rewritten as:
\[
\begin{split}
& EvalCmp^{Hope}(c_0, c_1) \\
=\quad & \frac{\left(1 + n\eta_0\phi(n)^\zeta(m_0-m_1)\right) \left( \left( \frac{r_0}{r_1}\right)^{n\phi(n)} \right)^{\eta_0\phi(n)^{\zeta-1}} \text{ mod } n^2 - 1}{n}\\
&\times \left(\eta_1 \cdot \phi(n)^{-\zeta} \text{ mod } n \right) \text{ smod } n,
\end{split}
\]
Per Lemma~\ref{thm:lagrange}, we know $\left( \frac{r_0}{r_1}\right)^{n\phi(n)} = 1 \text{ mod } n^2$,
therefore the above equation can be further simplified as the following:
\[
\begin{split}
& EvalCmp^{Hope}(c_0, c_1) \\
=\quad & \frac{\left(1 + n\eta_0\phi(n)^\zeta(m_0-m_1)\right) \left( 1 \right)^{\eta_0\phi(n)^{\zeta-1}} \text{ mod } n^2 - 1}{n}\\
&\times \left(\eta_1 \cdot \phi(n)^{-\zeta} \text{ mod } n \right) \text{ smod } n\\
=\quad & \frac{\left(1 + n\eta_0\phi(n)^\zeta(m_0-m_1)\right) \text{ mod } n^2 - 1}{n}\\
&\times \left(\eta_1 \cdot \phi(n)^{-\zeta} \text{ mod } n \right) \text{ smod } n\\
=\quad & \frac{n\eta_0\phi(n)^\zeta(m_0-m_1) \text{ mod } n^2}{n} \left(\eta_1 \phi(n)^{-\zeta} \text{ mod } n \right) \text{ smod } n\\
=\quad & \left( \eta_0\phi(n)^\zeta(m_0-m_1) \text{ mod } n^2 \right) \left(\eta_1 \phi(n)^{-\zeta} \text{ mod } n \right) \text{ smod } n\\
=\quad & \left( \eta_0\eta_1\left(\phi(n)^\zeta\phi(n)^{-\zeta}\right)(m_0-m_1) \text{ mod } n \right) \text{ smod } n\\
=\quad & \eta_0\eta_1(m_0-m_1) \text{ smod } n.
\end{split}
\]
Note that both $\zeta_0$ and $\zeta_1$ are random positive integers in $\mathbb{Z}^*_{n}$;
therefore the sign of $EvalCmp^{Hope}(c_0, c_1)$ is the same as $Sgn(m_0-m_1)$.

\section{Security Analysis}
\label{sec:security}

Only because HOPE homomorphically recovers the sign of the subtraction between two plaintexts does not mean HOPE is secure as shown in Section~\S\ref{sec:sub_correctness};
we will demonstrate the security of HOPE in this section.
We will follow the formal reduction framework to prove the computational security of a scheme;
the framework consists of three components:
the assumptions, the threat model (which is a part of security definition), and the security proof.

\subsection{Assumptions}

The security of HOPE depends on the hardness of a computational problem called the \textit{n-th residue},
which is also the assumption used by Paillier~\cite{ppail_eurocrypt99}.
It is believed that the n-th residue problem is intractable.
Informally, the problem is to identify an element $g \in \mathbb{Z}^*_{n^2}$ whose $n$-th residue is equal to a given value $h \in \mathbb{Z}^*_{n^2}$:
\[
g^n = h \texttt{ mod } n^2.
\]
In some sense, this is a dual problem of the well-known Discrete Logarithm problem in a cyclic group:
Given a base value and the modulo result of an unknown number of exponentiation, it is computationally difficult to find the exponent of the base that satisfies the equation.
In fact, the decisional version of this problem is believed to be intractable: 
It is computationally infeasible to decide whether $g$ exists such that $g^n = h \texttt{ mod } n^2$.

We give the formal n-th residue problem as follows.
\begin{definition}[The n-th residue problem]
\label{def:nmod}
    Let $n = pq$ as defined before. Given $h \in \mathbb{Z}^*_{n^2}$, is there any $g \in \mathbb{Z}^*_{n^2}$ satisfying $g^n = h \texttt{ mod } n^2$?
\end{definition}
It should be note that in practice the cardinality of $\mathbb{Z}^*_{n^2}$ is huge (e.g., by selecting large $p$ and $q$ primes to reach some exponentiation of the security parameter, $\phi(n^2) = \mathcal{O}(2^\lambda)$) and it is infeasible to enumerate all the elements in the group.

\subsection{Threat Model}
\label{sec:sub_model}

We will adopt the chosen-plaintext attack (CPA) threat model for the external attackers:
The adversary $\mathcal{A}$ could obtain the ciphertexts of a polynomial number of arbitrary plaintexts. 
However, due to the nature of equality support in order-preserving encryption,
the adversary is not allowed to repeat the same plaintext $m$ in the following two scenarios: 
(i) Querying the ciphertext of $m$ through an oracle (which is a hypothetical function that is accessible to the adversary), and
(ii) Sending $m$ as part of the pair of messages to the challenger in the distinguishing experiment.
The above requirement is necessary because otherwise the experiment can be trivially won by the adversary:
$\mathcal{A}$ can simply call the $EvalCmp^{Hope}()$ function to check which message is encrypted by the challenger.

The comparison key $ck$ is not visible to the general public and is kept by the outsourced database $\mathcal{D}$.
We assume that $\mathcal{D}$ is \textit{semi-honest}:
$\mathcal{D}$ will follow the HOPE protocol and will not collude with third-party adversaries; 
however $\mathcal{D}$ may intentionally run some algorithms over the ciphertexts.
Therefore, the adversaries cannot launch a frequency-analysis attack;
however, $\mathcal{D}$ could do so just as is entailed by the plaintext $Sgn()$ function as well.
That is, if we expect the outsourced database to check ciphertext equality such as \texttt{group by}, we have to reveal the frequency of tuples to $\mathcal{D}$ anyways.

\subsection{Security Proof}

There are a few security properties we need to demonstrate:
(i) No party except for the data owner can infer $\phi(n)$ from $ck$ as long as $n$ is reasonably large;
(ii) The adversary $\mathcal{A}$ cannot distinguish a ciphertext encrypted by HOPE even after obtaining ciphertexts of a polynomial number of plaintexts; and
(iii) The database $\mathcal{D}$ cannot obtain extra information of the plaintext distribution with $ck$ in the ciphertext space comparing with the hypothetical scenario with $Sgn()$ in the plaintext space.

\paragraph{Secrecy of private key $\phi(n)$ from $ck$}
While it is well known that the private key $sk = \phi(n)$ is computationally infeasible to derive from the public key $pk = n$ assuming the problem of factoring $n$ into $p$ and $q$ is hard,
we need to show that no PPT function exists to map from $ck$ to $sk$.
To make our analysis easier to read, let $\varphi = \phi(n)$.
Therefore, $ck = \left( \eta_0\varphi^\zeta, \eta_1\varphi^{-\zeta} \right)$ and $sk = \varphi$.
We prove the secrecy of $sk$ in the following theorem.
\begin{theorem}
    There does not exist a PPT algorithm that takes $ck$ and outputs $sk$.
\end{theorem}
\begin{proof}
    Assume for the sake of contradiction that there exists a PPT algorithm $\mathcal{A}$ satisfying $\mathcal{A}(ck) = sk$.
    This implies that given any $\eta_0$, $\mathcal{A}$ could recover $\varphi$.
    Let $\eta_0 = 1$ and $\zeta = n$, which makes it an easier problem, and we know that $\mathcal{A}\left( \varphi^n \right) = \varphi$.
    That is, $\mathcal{A}$ can efficiently compute the base of a given modulo value raised to the $n$-th power.
    Given the above computational capability, $\mathcal{A}$ can surely decide whether a given value can be written in the form of $\varphi^n$.
    However, this is exactly the $n$-th modulo problem as stated in Definition~\ref{def:nmod},
    which is believed intractable, leading to a contradiction.

    The case for $\eta_1$ is similar and we skip the detail.
\end{proof}

\paragraph{Semantic security from adversaries}

We will use the standard indistinguishability (IND) experiment to demonstrate the security of HOPE under the chosen-plaintext attack (CPA).
Note that the comparison key $ck$ is not visible to the adversary $\mathcal{A}$ in our threat model, as discussed in Section~\S\ref{sec:sub_model}.
Therefore, the algorithm $EvalSign^{Hope}$ is inaccessible to $\mathcal{A}$.
We formalize the property using the following theorem (recall that we ignore the key generation algorithm since it is trivial).
\begin{theorem}
    The scheme defined in the following quadruple 
    \[
    \left( Enc^{Hope}, Dec^{Hope}, EvalAdd^{Hope}, EvalSub^{Hope} \right)
    \]
    is IND-CPA secure.
\end{theorem}
\begin{proof}
    Let $m_0$ and $m_1$ denote two messages selected by $\mathcal{A}$.
    $\mathcal{A}$ sends both messages to a challenger $\mathcal{C}$,
    which encrypts a random message $m_b$, $b \in \{0, 1\}$:
    \[
    c \gets Enc^{Hope}(m_b).
    \]
    $\mathcal{C}$ sends $c$ back to $\mathcal{A}$, and let us assume that $\mathcal{A}$ as a function can do the following:
    \[
    \mathcal{A}(c) = b,
    \]
    with a probability larger than $\frac{1}{2}$ by a non-negligible function in $n$.
    That is, we assume that HOPE is insecure.
    This means that $\mathcal{A}$ can tell the difference between $c$ and a random string, say $str$, by a probability at least $\frac{1}{2} + \frac{1}{Poly(n)}$.
    Recall that 
    \[
    c \gets (n + 1)^{m_b \text{ mod } n} \cdot r^n \text{ mod } n^2,
    \]
    as defined in Section~\S\ref{sec:sub_smod}.
    Now, we consider a simpler problem, where $m \in (0, n)$ and $r = 1$,
    implying that $c' = (n+1)^{m_b}$.
    If $\mathcal{A}$ can tell the difference between $c$ and $str$,
    $\mathcal{A}$ can certainly distinguish between $(n+1)^{m_b} \text{ mod } n^2$ and $str$.
    That is, $\mathcal{A}$ can decide whether $m_b$ is the $(n+1)$-based discrete logarithm of a given $c'$ with a probability at least $\frac{1}{2} + \frac{1}{Poly(n)}$, significantly better than a random guess.
    However, this is impossible because the (decision variant of) discrete logarithm problem is intractable.
    Therefore, we reach a contradiction and prove the claim.
\end{proof}

\paragraph{Equivalent semantics for outsourced databases}
We will show that the database $\mathcal{D}$ with the $EvalSign^{Hope}()$ function in the ciphertext space gains no extra information regarding the plaintext distribution compared to with the $Sgn()$ function in the plaintext space.
We will refer to combination of ciphertext space and $EvalSign^{Hope}$ as the \textit{real world} and the combination of plaintext space and $Sgn()$ as the \textit{ideal world}.
We will prove the following claim.
\begin{theorem}
All functions in the real world can be simulated in the ideal world.
\end{theorem}
\begin{proof}
    The proof will take the \textit{simulation} framework in cryptography to demonstrate that any effects in the real world can be similarly made in the ideal world;
    the rationale is that if such a simulation is always possible, then the real world leaks no more information than the ideal world that is aligned with the design goal and meets the application requirements.
    In other worlds, this simulation ensures that all the behaviors in the real world are ``allowed'' because they are possible in the ideal world.
    In the proposed HOPE scheme, the sign information in the ciphertext space is exposed through the $EvalCmp^{Hope}(c_0, c_1)$ algorithm that outputs $\eta_0\eta_1(m_0 - m_1)$,
    which is obviously not more than the output of $Sgn(m_0 - m_1)$ in the plaintext space.
    Therefore, we need to show that $\eta_0\eta_1(m_0 - m_1)$ can be \textit{simulated} by $Sgn(m_0 - m_1)$.
    Intuitively speaking, we want to show that $\eta_0\eta_1(m_0 - m_1)$ does not leak $(m_0 - m_1)$ with a probability up to a negligible function in $n$.
    Recall that $\eta_0$ and $\eta_1$ are sampled from $\mathbb{Z}^*_n$;
    therefore $\eta = \eta_0\eta_1 \in \mathbb{Z}^*_n$ in the multiplicative group.
    In other worlds, it is always possible to pick $\eta$ and multiply it with $Sgn(m_0 - m_1)$ in the plaintext space to simulate the value of $EvalCmp^{Hope}(c_0, c_1)$.
    The probability of $\eta(m_0 - m_1) = (m_0 - m_1)$ is obviously on the order of the inverse of an exponential function in $n$, which is a negligible function.
\end{proof}

\section{Conclusion and Future Work}

This paper proposes a new OPE scheme,
namely homomorphic order-preserving encryption (HOPE),
which does not require client storage and does not incur any client-server interaction during the queries.
The key idea of HOPE is to leverage the underlying additive property of a homomorphic encryption scheme such that the sign of the difference between two plaintexts can be revealed by some algebraic operations with an evaluation key that is accessible to the database server for only revealing the order of ciphertexts,
which remain randomized for the adversary.
The paper details the design of HOPE and analyzes its correctness and feasibility.
In addition, the security of HOPE is formally proved using standard cryptographic frameworks.

It is our ongoing effort to implement HOPE in MySQL loadable functions using C++.
If there is enough interest, we also plan to implement it as a PostgreSQL extension using C.
We expect to evaluate the HOPE-MySQL integration using multiple standard micro-benchmarks including TPC-H as well as real-world data sets on outsourced database instances deployed on both industry and government-sponsored cloud platforms, such as Amazon Web Services and the Chameleon Cloud.


{\footnotesize
\bibliographystyle{acm}
\bibliography{ref_new}
}

\end{document}